\documentclass{article}
\usepackage[utf8]{inputenc}
\usepackage[english]{babel}
\usepackage{color}
\usepackage{amsmath}
\usepackage{amssymb}
\usepackage{amsthm}
\usepackage{graphicx}
\usepackage{url}
\usepackage{natbib}
\title{Comment on \cite{Wibral2013}: Measuring Information-Transfer Delays \\
(Correction of the previous comment)
}
\author{Jakob Runge\\
{\small Potsdam Institute for Climate Impact Research (PIK), 14473 Potsdam, Germany}\\
{\small Department of Physics, Humboldt University, 12489 Berlin, Germany}
}
\date{\today}

\begin{document}
%[As anadvancement to Pompe2011, in Runge2012ab we propose the approach that first the causal links are estimated with the same measure as WIbral, and in a second step \emph{only for the causal links} the ``weights'' of the links are estimated using MIT...]

\maketitle
In their article \cite{Wibral2013}, the authors propose a measure of interaction delays rooted in an information-theoretic framework. Their measure, named $TE_{SPO}$ where $SPO$ stands for \emph{self-prediction optimality}, is a time-delayed extension to transfer entropy that becomes maximal only at the actual interaction delay as proven in their paper. 

Wibral et al. only considered the bivariate coupling case. Prior to their publication, in \cite{Runge2012prl,Runge2012b} the theory of detecting and quantifying causal interactions and their strength for the general multivariate case is discussed. In these publications we use a measure, called information tranfer to Y (ITY), that can be considered as a multivariate generalization of $TE_{SPO}$. The general idea is to measure the coupling delay $X\to Y$ by excluding past information from $Y$. The difference is that Wibral et al. try to infer the past information by reconstructing (vector-valued) states, while in our approach (published earlier than theirs) we use an iterative algorithm coming from the statistics community \citep{Pearl2000}.
%However, Wibral et al. present an interesting complementary approach in that they propose to determine the interaction delay based on the reconstructed (vector-valued) states rather than the (scalar) observations of the complex systems under study.

Wibral et al. contrast their measure with a similar information-theoretic measure, the bivariate momentary information transfer (BivMIT), that was introduced before the two main publications \citep{Runge2012prl,Runge2012b} in \cite{Pompe2011}.
Wibral et al. write that
``\emph{A major conceptual difference between the Pompe and Runge
study and ours is that no formal proof of the maximality of their
functional [Biv-]MIT at the correct interaction delay is given, and as we
argue below cannot be given.}'' 
Wibral et al. do not provide a proof that the maximality of BivMIT cannot be given, but they construct a model example for which they find the BivMIT unable to serve the purpose of inferring the interaction delay. 
In a previous version of this comment we had attemped to show that the maximality proof equally holds for the bivariate BivMIT introduced in  \cite{Pompe2011}. But this proof contained an error as Wibral et al. brought to our attention. Here we show, that indeed, the BivMIT could be maximal for a different delay. Interestingly, BivMIT can be regarded to some extent as a derivative of $TE_{SPO}$ and the maximum, therefore, depends on the decay of $TE_{SPO}$. 

While the maximality of $TE_{SPO}$ can be proven for the bivariate case with \emph{one} coupling delay (possibly also in the other direction $Y\to X$), $TE_{SPO}$ is not necessarily maximal anymore if there are \emph{two} coupling delays from $X$ to $Y$. These more complicated cases are only captured by the iterative approach introduced in \cite{Runge2012prl,Runge2012b}. In these works a two-step approach is proposed. In the first step the causal graph (conditional independence graph) is iteratively reconstructed \citep{Runge2012prl} using the well-established framework of graphical models \citep{Pearl2000}, where the property to capture the correct causal interaction delays is a consequence of separation properties in the graph. In \cite{Runge2012prl} also the underlying assumptions for such an inference are given. In the second step the multivariate MIT (not the bivariate BivMIT introduced in \cite{Pompe2011}) is used as a measure of the coupling strength solely of the causal links, i.e., the inferred interaction delays, in this graph \citep{Runge2012b}.

Wibral at al. also have overseen that their estimator of conditional mutual information given by their Eq.~(19) has already been developed in \cite{FrenzelPompe2007}. The latter work also discussed the inference of interaction delays.

\section*{Proof that BivMIT is not necessarily maximal at the correct interaction delay and $TE_{SPO}$ not maximal anymore for two coupling delays}

BivMIT is defined as
\begin{align} \label{eq:def_mit}
 I^{\rm BivMIT}_{X{\to}Y}(\delta) &\equiv I(X_{t-\delta};Y_t| Y^-_t, X^-_{t-\delta}),
\end{align}
where $Y^-_t=(Y_{t-1},\ldots),\,X^-_{t-\delta}=(X_{t-\delta-1},\ldots)$ are the embedding vectors of the past state. I.e., additionally to $TE_{SPO}$, the past state of $X$ is included in the condition. 

Wibral et al.  prove that their $TE_{SPO}$ is maximal only for the interaction delay. To this end they use the Markov properties of the process and information theoretic properties of the conditional mutual information like the data processing inequality and the chain rule \citep{Cover2006} which we also utilize.

First, consider the unidirectional case with only one delay $X\to Y$ at lag $\tau$. Then $I^{\rm BivMIT}_{X{\to}Y}(\delta)=0$ for $\delta>\tau$ because the condition on $X^-_{t-\delta}=(X_{t-\delta-1},\ldots)$ excludes all paths coming from $X_{t-\tau}$. On the other hand $TE_{SPO}$ is not zero for smaller lags. In this way BivMIT could be used to measure the coupling delay by checking at which lag it becomes non-zero, but it only holds for unidirectional couplings.

Turning to the more general case, to proof that BivMIT is not necessarily maximal at the correct coupling delay, we simplify the notation by leaving out the time index $t$ in the subscripts, i.e., $X^-=(X_t, X_{t-1},\ldots),~X_\delta=X_{t-\delta}$ and so on. We assume that there is only one coupling delay at $\tau$ (possibly also in the other direction) and the past state is only the preceding time point, i.e., $X^-_{t-\delta}=(X_{t-\delta-1})$.
Now consider the multivariate conditional mutual information with the chain rule applied twice extracting first $X_{\delta}$ and then $X_{\delta+1}$:
\begin{align}
& I(X^-;Y | Y^-)= \nonumber\\
& =I(X_\delta;Y|Y^-) + I(X^-\setminus X_\delta;Y|Y^-,X_\delta) \\
& = I(X_\delta;Y|Y^-) + I(X_{\delta+1};Y|Y^-,X_\delta) + I(X^-\setminus (X_\delta,X_{\delta+1});Y|Y^-,X_\delta,X_{\delta+1}) \label{one}.
\end{align}
On the other hand, we can also extract the vector $(X_{\delta},X_{\delta+1})$ at once and then extract $X_{\delta+1}$ and then $X_{\delta}$:
\begin{align}
& I(X^-;Y | Y^-)= \nonumber\\
& =I((X_{\delta},X_{\delta+1});Y|Y^-) + I(X^-\setminus (X_{\delta},X_{\delta+1});Y|Y^-,X_{\delta},X_{\delta+1}) \\
& = I(X_{\delta+1};Y|Y^-) + I(X_\delta;Y|Y^-,X_{\delta+1}) + I(X^-\setminus (X_\delta,X_{\delta+1});Y|Y^-,X_\delta,X_{\delta+1}) \label{two}.
\end{align}
Now the last terms in Eqns.~(\ref{one}) and (\ref{two}) cancel and the BivMIT at lag $\delta$ and the true coupling delay $\tau$ are
\begin{align}
& I(X_\delta;Y|Y^-,X_{\delta+1}) = I(X_\delta;Y|Y^-) - I(X_{\delta+1};Y|Y^-) +  I(X_{\delta+1};Y|Y^-,X_\delta) \\
& I(X_\tau;Y|Y^-,X_{\tau+1}) = I(X_\tau;Y|Y^-) - I(X_{\tau+1};Y|Y^-) +  \underbrace{I(X_{\tau+1};Y|Y^-,X_\tau) }_{\text{$=0$ due to the condition on $X_\tau$}}.
\end{align}
This implies, that BivMIT at $\tau$ is the difference of $TE_{SPO}$ at lags $\tau+1$ and $\tau$ and analogously for $\delta$ where another term is added. This can be regarded as a derivative and depending on the decay rate of $TE_{SPO}$ with $\delta$ the maximum shifts. Here, we don't examine further for what kind of processes the maximum shifts, Wibral et al. have given one example. It has a \emph{deterministic} coupling from $X$ to $Y$ and randomness only in $X$. If even this remaining randomness is set to very small values, only then does the ``inability'' of MIT appear. Generally, information theory is primarily suited for stochastic processes. As discussed also in the original paper \citep{Pompe2011}, deterministic processes generate the source entropy that MIT is based on only in the chaotic case due to quantization effects. 

Nevertheless, it cannot be proven for the class of processes sufficing condition (S) in \cite{Eichler2011} that BivMIT reliably infers the correct coupling delay and, as analyzed in detail in the subsequent articles \cite{Runge2012prl} and \cite{Runge2012b}, we propose the iterative approch for the general, also multivariate, case.

To show that $TE_{SPO}$ fails if there are two couplings $\tau_1, ~\tau_2$ in one direction, consider 
\begin{align}
& I((X_{\delta},X_{\tau_1},X_{\tau_2});Y | Y^-)= \nonumber\\
& = I(X_{\delta};Y|Y^-) +  \underbrace{I((X_{\tau_1}, X_{\tau_2});Y|Y^-,X_\delta)}_{\geq 0} \\
& = I((X_{\tau_1}, X_{\tau_2});Y|Y^-) + \underbrace{I(X_{\delta};Y|Y^-,X_{\tau_1}, X_{\tau_2})}_{\text{$=0$ due to the condition on $X_{\tau_1}$ and $X_{\tau_2}$}}\\
& = I(X_{\tau_1};Y|Y^-) + I(X_{\tau_2};Y|Y^-,X_{\tau_1})\\
& (= I(X_{\tau_2};Y|Y^-) + I(X_{\tau_1};Y|Y^-,X_{\tau_2})).
\end{align}
Therefore, $TE_{SPO}$ given by $I(X_{\delta};Y|Y^-)$ can be proven to be smaller than the multivariate $I((X_{\tau_1}, X_{\tau_2});Y|Y^-)$, but not necessarily smaller than each individual $I(X_{\tau_1};Y|Y^-))$. Again, this more general case requires an iterative approach.

\end{document}